\documentclass[12pt,a4paper]{article}

\usepackage[utf8]{inputenc}
\usepackage[T1]{fontenc}
\usepackage[margin=2.5cm]{geometry}
\usepackage{graphicx}
\usepackage{booktabs}
\usepackage{amsmath}
\usepackage{hyperref}
\usepackage{float}
\usepackage{setspace}
\usepackage[numbers,sort&compress]{natbib}
\usepackage{caption}
\usepackage{subcaption}
\usepackage{url}
\hypersetup{
    colorlinks=true,
    linkcolor=blue,
    citecolor=blue,
    urlcolor=blue
}

\title{\textbf{DR.~INFO at the Point of Care: A Prospective Pilot Study of Physician-Perceived Value of an Agentic AI Clinical Assistant}}

\author{
Rog\'{e}rio Corga da Silva$^{1,2}$, Miguel Romano$^{1,3,4,5}$, Tiago Mendes$^{1,6}$, \\
Marta Isidoro$^{1,7,8}$, Sandhanakrishnan Ravichandran$^{1}$, Shivesh Kumar$^{1}$, \\
Michiel van der Heijden$^{1}$, Olivier Fail$^{1}$, \\
Valentine Emmanuel Gnanapragasam$^{1}$\thanks{Corresponding author: valentine.emmanuel@synduct.com} \\[6pt]
\small{$^{1}$DR.~INFO, Synduct GmbH, Munich, DEU} \\
\small{$^{2}$Critical Care, Hospital de Santa Luzia, Viana do Castelo, PRT} \\
\small{$^{3}$Medical Education and Simulation, SIMBOX, Viana do Castelo, PRT} \\
\small{$^{4}$Intensive Care and Internal Medicine, Unidade Local de Sa\'{u}de do Alto Minho, Viana do Castelo, PRT} \\
\small{$^{5}$School of Medicine, University of Minho, Braga, PRT} \\
\small{$^{6}$Medicine, Unidade Local de Sa\'{u}de do Alto Minho, Ponte de Lima, PRT} \\
\small{$^{7}$Anaesthesiology, Unidade Local de Sa\'{u}de de Almada-Seixal, Lisbon, PRT} \\
\small{$^{8}$Anaesthesiology, Faculdade de Ci\^{e}ncias M\'{e}dicas de Lisboa, Lisbon, PRT}
}

\date{}

\begin{document}
\maketitle


\begin{abstract}

\textbf{Background:} Clinical documentation and information retrieval consume over half of physicians' working hours, contributing to cognitive overload and burnout. While artificial intelligence (AI) offers a potential solution, concerns over hallucinations and source reliability have limited adoption at the point of care. This study aimed to evaluate physician-perceived time efficiency, decision-making support, and satisfaction with DR.~INFO, an agentic AI clinical assistant, in routine clinical practice.

\textbf{Methodology:} In this prospective, single-arm, pilot feasibility study, 29 physicians and medical students across multiple specialties in Portuguese healthcare institutions used DR.~INFO v1.0 over five working days within a two-week period. Outcomes were assessed via daily Likert-scale evaluations (time saving and decision support) and a final Net Promoter Score (NPS). Non-parametric methods were used throughout, with bootstrap confidence intervals (CIs) and sensitivity analysis to address non-response.

\textbf{Results:} Physicians reported high perceived time saving (mean = 4.27/5; 95\% CI = 3.97--4.57) and decision support (mean = 4.16/5; 95\% CI = 3.86--4.45), with ratings stable across the five-day study window. Among the 16 (55\%) participants who completed the final evaluation, the NPS was 81.2, with no detractors; sensitivity analysis indicated an NPS of 44.8 under conservative non-response assumptions.

\textbf{Conclusions:} Physicians across specialties and career stages reported positive perceptions of DR.~INFO for both time efficiency and clinical decision support within the study window. These findings are preliminary and should be confirmed in larger, controlled studies that include objective performance measures and independent accuracy verification.

\vspace{6pt}

\end{abstract}

\newpage


\section{Introduction}

The volume of biomedical literature and clinical data generated daily far exceeds what any individual physician can assimilate, widening the gap between available evidence and bedside practice \cite{liu2025}. According to the American Medical Association, physicians work an average of 57.8 hours per week, of which only 27.2 are spent in direct patient care \cite{berg2025}. The balance is consumed by electronic health record documentation and administrative tasks, contributing to the 43.2\% burnout prevalence reported nationally \cite{berg2025,shanafelt2025}.

Artificial intelligence (AI) has been proposed as a means of reducing this burden \cite{hadly2025}. Recent studies have demonstrated that large language models (LLMs) can match or exceed physician performance on standardized medical examinations and patient queries \cite{singhal2025,ayers2023,thirunavukarasu2023}, and a randomized clinical trial showed that access to an LLM influenced physician diagnostic reasoning \cite{goh2024}. Early usability assessments have also explored potential roles for conversational AI across the clinical workflow \cite{rao2023}. However, the clinical adoption of general-purpose LLMs has been limited by well-documented safety concerns, most notably hallucinations, defined as the generation of plausible but factually incorrect outputs \cite{omar2025}. These evaluations have primarily tested standalone models rather than integrated clinical assistants deployed in routine practice.

DR.~INFO is an agentic AI clinical assistant designed to reduce these risks by coupling LLM capabilities with retrieval from curated medical knowledge bases and peer-reviewed literature. In the context of this study, the term agentic refers to a system that executes multi-step clinical information workflows: it decomposes a physician's query, performs clinical reasoning, retrieves and verifies sources from curated medical databases, and synthesizes a cited response. DR.~INFO does not access electronic health records, generate prescriptions, or perform autonomous clinical actions. The physician retains full decision-making authority at all times. Between 2024 and 2025, Portugal ranked first in the European Union (EU) for access to digital health data \cite{compete2030}, providing a favorable environment for the deployment of DR.~INFO across several Local Health Units and private hospitals. This deployment coincides with the enforcement of the EU AI Act, which requires transparency and human oversight for high-risk AI systems. These principles guided the design of DR.~INFO \cite{spms2025}.

Before this pilot, DR.~INFO was developed through a user-centered co-design process involving a multi-specialty group of physicians, referred to as the ``Innovator Circle,'' who provided iterative feedback during development. This group was not financially compensated and had no formal affiliation with the study team. This pilot enrolled physicians and medical students at the participating institutions independently of prior Innovator Circle involvement.

\subsection{Technical benchmarking: HealthBench}

Separately from this usability study, the technical performance of DR.~INFO was evaluated using OpenAI's HealthBench, a benchmark comprising 5,000 real-world clinical conversations scored across the following five dimensions: accuracy, completeness, context awareness, communication quality, and instruction following \cite{openai2025}. The methodology and results are reported in a companion paper \cite{ravichandran2025}. These comparisons reflect differences in system architecture: DR.~INFO uses an agentic architecture that retrieves from curated medical knowledge bases and peer-reviewed literature before generating a response, whereas comparator models were evaluated as standalone LLMs without a clinical retrieval infrastructure. The scores, therefore, reflect the combined value of architecture and retrieval, not model quality alone.

On the 1,000-case ``Hard'' subset, DR.~INFO achieved a score of 0.68, compared with GPT-5 (0.40--0.46), Grok 3 (0.23), and Gemini 2.5 Pro (0.19). The standalone LLM scores were sourced from the HealthBench evaluation conducted by OpenAI (May 2025) \cite{openai2025}. Against clinical AI assistants with similar agentic capabilities, DR.~INFO scored 0.72 versus OpenEvidence (0.49) and DoxGPT (0.48). Full comparison details are provided in the companion study \cite{ravichandran2025}. Given the pace of model development, these benchmark scores reflect performance at the time of evaluation and are expected to evolve as newer model versions are released.

In this pilot study, we assess physician-perceived time efficiency, decision-making support, and satisfaction with DR.~INFO in routine clinical practice. The primary objective was to quantify physician-perceived time saving on a five-point Likert scale. Secondary objectives were to assess physician-perceived decision-making support on a matching Likert scale and overall satisfaction through the Net Promoter Score (NPS). As a pilot feasibility study consistent with Thabane et al. \cite{thabane2010}, no formal hypothesis was tested. All findings are exploratory and intended to guide future controlled evaluations.


\section{Materials and Methods}

\subsection{Pilot study design}

This prospective, single-arm pilot feasibility study \cite{thabane2010} was conducted over a two-week period in October 2025 using DR.~INFO v1.0. Eligibility was defined as active clinical practice (physicians and medical students) at the participating institutions during the study window; no exclusion criteria were applied. The sample of 29 reflects all available participants and represents a convenience cohort. No formal power calculation was performed, as the sample size in pilot feasibility studies is guided by practical considerations rather than statistical inference \cite{thabane2010}. Each participant used DR.~INFO for five working days during the two-week study window, selected based on their clinical availability.

DR.~INFO is an agentic, retrieval-augmented generation system that combines an LLM with curated medical knowledge bases and peer-reviewed literature. Physician queries are processed through multi-step workflows that include decomposition, clinical reasoning, source retrieval and verification, and synthesis of cited responses. Physicians accessed the tool via a conversational interface on their personal devices (smartphones and computers), operating independently of electronic health record systems. Full technical specifications are reported in the companion evaluation \cite{ravichandran2025}.

In total, 29 participants were enrolled on a voluntary basis, comprising 25 physicians and four medical students. Of these, 26 completed the baseline electronic Case Report Form (eCRF), 28 submitted at least one diary entry (90 named entries total), and 16 completed the final evaluation. Three participants completed the final evaluation without having submitted the baseline form, and 13 did not complete the final evaluation.

\subsection{Instruments and protocol}

Three eCRFs were administered in sequence. The baseline form (eCRF 1) captured years of practice as an ordinal variable ($<$1, $<$3, $<$10, $>$10 years), specialty, frequency of clinical decision-support tool usage (``How frequently do you use clinical decision support tools in your practice?''), and technology comfort rated on a five-point Likert scale (``How comfortable are you with technology, e.g., smartphones, AI, new software?'').

The daily form (eCRF 2) logged two primary outcome measures on five-point Likert scales \cite{sullivan2013} (1 = strongly disagree, 5 = strongly agree): physician-perceived time saving (``To what extent do you agree that using DR.~INFO today saved you time?'') and physician-perceived decision support (``To what extent do you agree that DR.~INFO helped you improve your clinical decisions today, e.g., information seeking, information retrieval, diagnosis, treatment, management?''). Participants also selected applicable clinical use cases from a predefined list and provided optional qualitative examples.

The final form (eCRF 3) collected the NPS \cite{reichheld2003} (``On a scale of 0 to 10, how likely are you to recommend DR.~INFO to a colleague?'') and two open-ended qualitative questions: ``What did you like most about DR.~INFO?'' and ``What did you like least about DR.~INFO? What could be improved?''

\subsection{Statistical analysis}

Given the exploratory nature of the study, the ordinal measurement scale, and the small sample size, non-parametric methods were used throughout. Attrition analysis compared diary scores between final eCRF completers and non-completers using the Mann-Whitney U test. In this study, 95\% confidence intervals (CIs) were computed using both percentile bootstrap resampling (10,000 iterations, resampled at the participant level) and t-distribution methods. Longitudinal stability within the study window was assessed using the Friedman test (non-parametric repeated measures) applied to the 10 participants with data on all five study days. Spearman's rank correlation coefficients were calculated for associations between ordinal variables (years of practice, technology comfort, and tool usage frequency) and the two primary outcome measures (time saving and decision-support scores).

Role-based data are presented descriptively in the Results section, given that subgroup sizes (n = 1 to n = 12) were below the threshold for reliable inferential testing \cite{siegel1988}. Formal statistical comparison across roles was, therefore, not performed. Because 13 of 29 participants did not complete the final eCRF, NPS was recalculated under alternative non-response assumptions (e.g., all non-responders scored as Passives, all as Detractors) to examine the sensitivity of the observed result to non-response bias. Open-ended responses were categorized into recurring themes (e.g., speed, source transparency, interface) through content analysis of the 16 responses by a single coder.

All analyses were conducted at the participant level (using each participant's mean across diary entries) to avoid pseudoreplication from repeated measures. No correction for multiple comparisons was applied. All reported associations are exploratory and hypothesis-generating.

\subsection{Data quality and transparency}

During data cleaning, several issues were identified and corrected. Free-text specialty entries were standardized from 24 unique labels to seven categories. Name inconsistencies across forms, such as abbreviated entries on later forms, were resolved through manual matching, reducing the apparent 31 unique diary names to 29 true participants. Four anonymous diary entries were excluded from participant-level analyses. Future studies should implement unique participant identifiers assigned at enrolment to eliminate the need for manual reconciliation.

\subsection{Ethical considerations}

This study involved voluntary participation of physicians and medical students who provided self-reported feedback through electronic questionnaires and CRFs. No patients were involved, and no patient data were collected or processed. Given the non-interventional design and minimal risk nature of the study, formal ethics committee approval was not required under Lei n.\textsuperscript{o} 21/2014 \cite{lei2014}. All participants provided informed consent before enrolment. Participation was voluntary, and participants could withdraw at any time without consequence. No incentives were offered.


\section{Results}

\begin{figure}[H]
    \centering
    \includegraphics[width=0.7\textwidth]{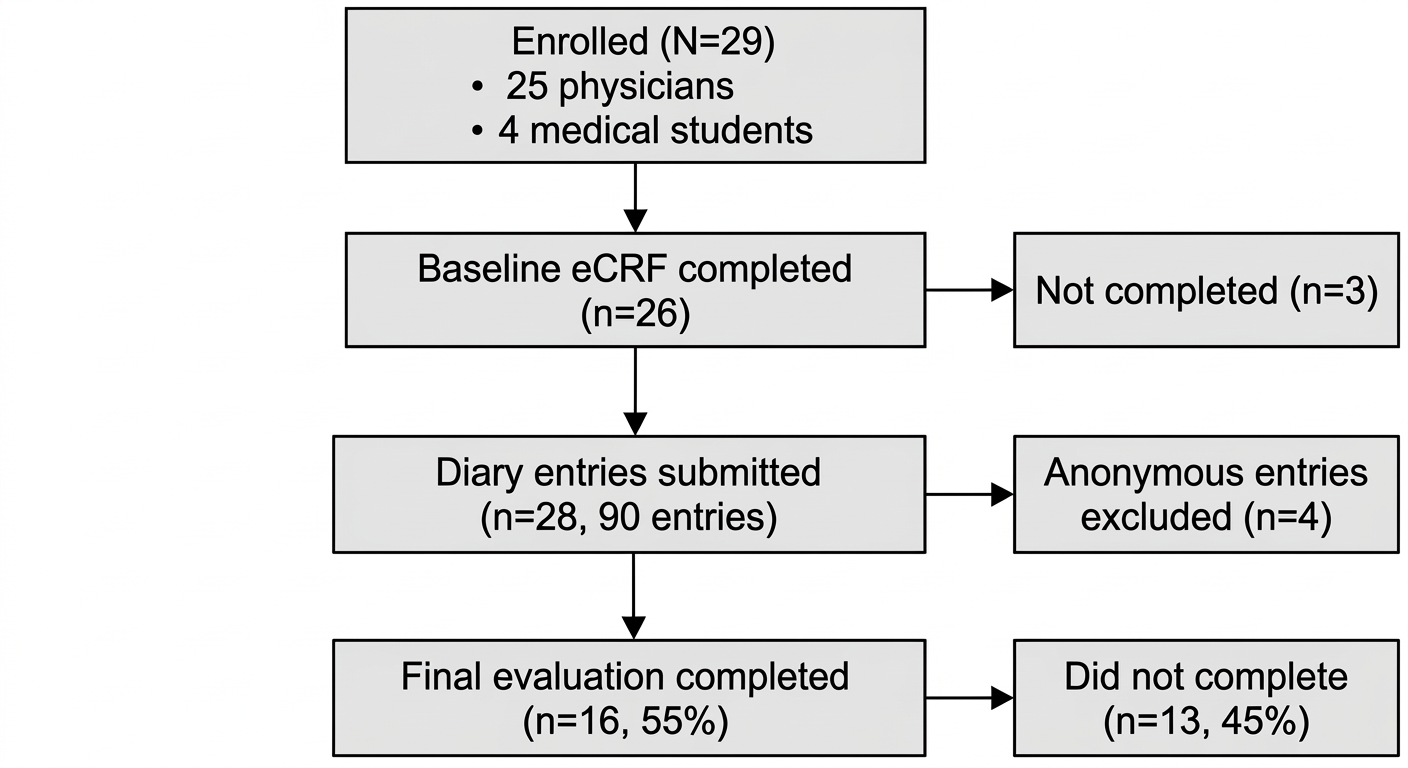}
    \caption{Participant flow diagram. Flow diagram showing participant progression through the study. Of the 29 enrolled participants (25 physicians and four medical students), 26 completed the baseline eCRF, 28 submitted at least one daily diary entry (90 named entries total, with four anonymous entries excluded), and 16 (55\%) completed the final NPS evaluation. Three participants did not complete the baseline form, and 13 did not complete the final evaluation. eCRF = electronic Case Report Form; NPS = Net Promoter Score.}
    \label{fig:consort}
\end{figure}

A participant flow diagram (Figure~\ref{fig:consort}) details the progression from enrolment through each data collection phase.

Of the 26 participants with role information, 85\% (n = 22) were practising physicians across Family Medicine (35\%), Intensive Care Medicine (23\%), Internal Medicine (12\%), Surgery, Cardiology, Anesthesiology, and Emergency Medicine. The remaining 15\% (n = 4) were medical students. The distribution of years of practice is shown in Table~\ref{tab:demographics}.

\begin{table}[H]
\centering
\caption{Participant demographics and experience distribution. N = 26 participants with baseline data. Three participants did not complete the baseline eCRF. eCRF = electronic Case Report Form.}
\label{tab:demographics}
\begin{tabular}{lcc}
\toprule
\textbf{Years of practice (category)} & \textbf{n} & \textbf{Percentage} \\
\midrule
$<$1 year & 4 & 15\% \\
1 to 3 years & 5 & 19\% \\
3 to 10 years & 7 & 27\% \\
$>$10 years & 10 & 38\% \\
\midrule
\textit{Missing} & \textit{3} & -- \\
\bottomrule
\end{tabular}
\end{table}

Baseline data from the first eCRF indicated that participants were already regular users of clinical decision-support tools: 36\% reported daily use, 60\% several times per week, and 4\% once per week. No participant reported using such tools less than once a month.

\subsection{Attrition analysis}
\label{sec:attrition}

\begin{figure}[H]
    \centering
    \includegraphics[width=0.7\textwidth]{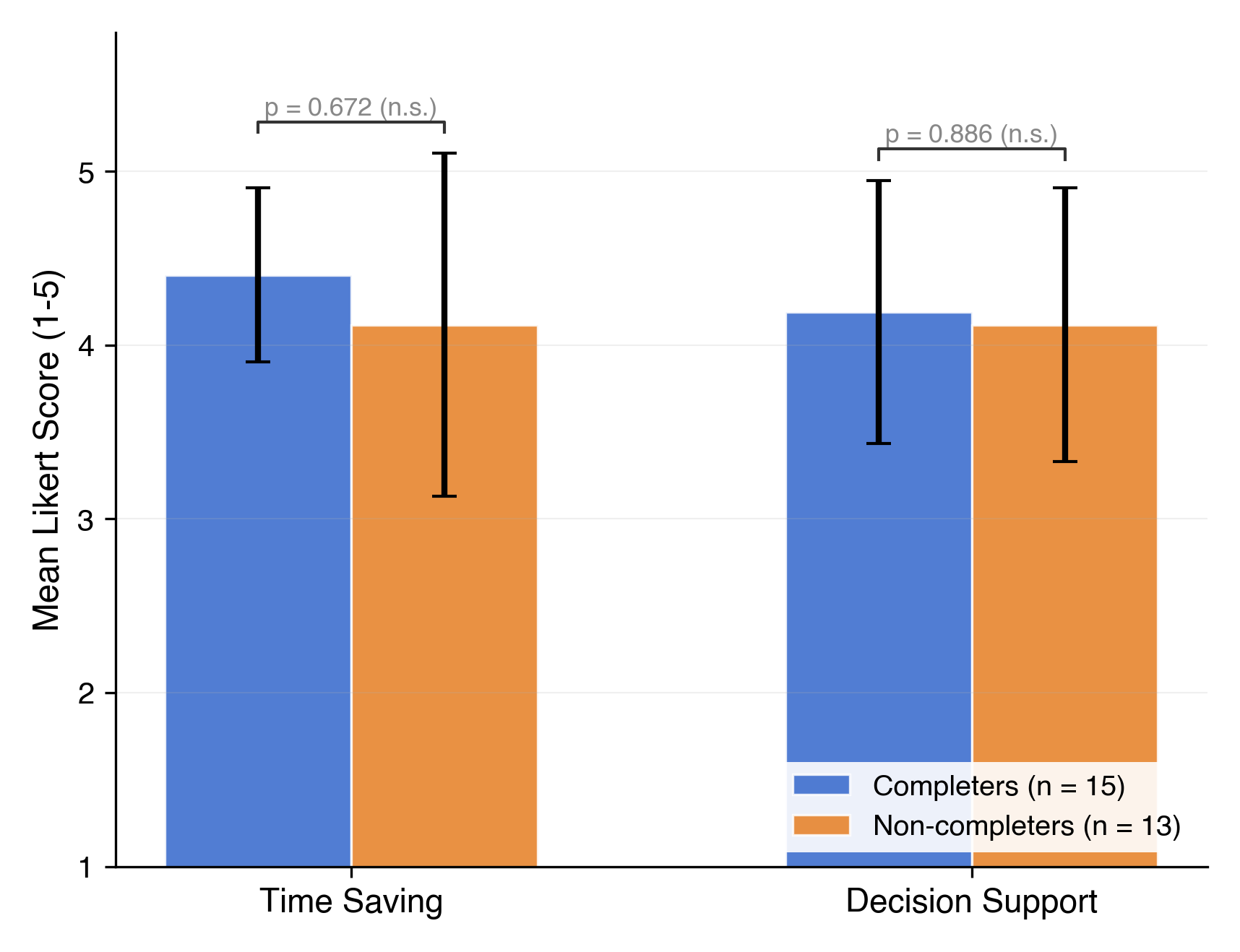}
    \caption{Comparison of mean diary scores between final eCRF completers and non-completers. Bar chart comparing participant-level mean Likert scores for time saving and decision support between those who completed the final evaluation (n = 15) and those who did not (n = 13). Error bars represent standard deviation. The Mann-Whitney U test did not detect statistically significant differences between the two groups for either time saving (U = 107.0, p = 0.672) or decision support (U = 101.0, p = 0.886), though the small sample size limits the ability to detect moderate differences. eCRF = electronic Case Report Form; n.s. = not significant.}
    \label{fig:attrition}
\end{figure}

Of the 29 enrolled participants, 16 (55\%) completed the final eCRF. To assess whether this attrition introduced systematic bias, we compared diary scores between final form completers (n = 15) and non-completers (n = 13) among the 28 participants who submitted at least one diary entry. As shown in Table~\ref{tab:attrition} and Figure~\ref{fig:attrition}, the Mann-Whitney U test did not detect statistically significant differences between the two groups on either metric (p = 0.672 for time saving, p = 0.886 for decision support), though the comparison was underpowered to detect moderate effects.

\begin{table}[H]
\centering
\caption{Attrition analysis: comparison of diary scores between final eCRF completers and non-completers. N = 28 participants with at least one diary entry. Groups were compared using the Mann-Whitney U test. eCRF = electronic Case Report Form; SD = standard deviation; U = Mann-Whitney U statistic.}
\label{tab:attrition}
\small
\begin{tabular}{lcccc}
\toprule
\textbf{Metric} & \textbf{Completers (n = 15)} & \textbf{Non-completers (n = 13)} & \textbf{U} & \textbf{P-value} \\
\midrule
Time saving (mean $\pm$ SD) & 4.40 $\pm$ 0.50 & 4.12 $\pm$ 0.99 & 107.0 & 0.672 \\
Decision support (mean $\pm$ SD) & 4.19 $\pm$ 0.76 & 4.12 $\pm$ 0.79 & 101.0 & 0.886 \\
\bottomrule
\end{tabular}
\end{table}

\subsection{Physician-perceived time savings and clinical reasoning support}

The participant-level mean score for physician-perceived time saving was 4.27 out of 5 (SD = 0.76; 95\% CI = 3.97--4.57), with 87.8\% of diary entries indicating agreement or strong agreement. Physician-perceived clinical decision support showed comparable results, with a mean of 4.16 (SD = 0.76; 95\% CI = 3.86--4.45) and 85.6\% agreement that the tool supported diagnostic or therapeutic reasoning. Table~\ref{tab:primary} presents the primary outcome metrics with 95\% CIs computed using both bootstrap resampling and t distribution methods.

\begin{table}[H]
\centering
\caption{Summary of primary outcome metrics with confidence intervals. N = 28 participants with at least one diary entry. Bootstrap CI computed using percentile resampling (10,000 iterations) at the participant level. Scores measured on a five-point Likert scale (1 = strongly disagree, 5 = strongly agree). SD = standard deviation; CI = confidence interval.}
\label{tab:primary}
\begin{tabular}{lcccc}
\toprule
\textbf{Metric} & \textbf{Mean} & \textbf{SD} & \textbf{95\% CI (bootstrap)} & \textbf{95\% CI (t based)} \\
\midrule
Time saving & 4.27 & 0.76 & 3.98--4.53 & 3.97--4.57 \\
Decision support & 4.16 & 0.76 & 3.87--4.43 & 3.86--4.45 \\
\bottomrule
\end{tabular}
\end{table}

\subsection{Longitudinal stability}

\begin{figure}[H]
    \centering
    \includegraphics[width=0.75\textwidth]{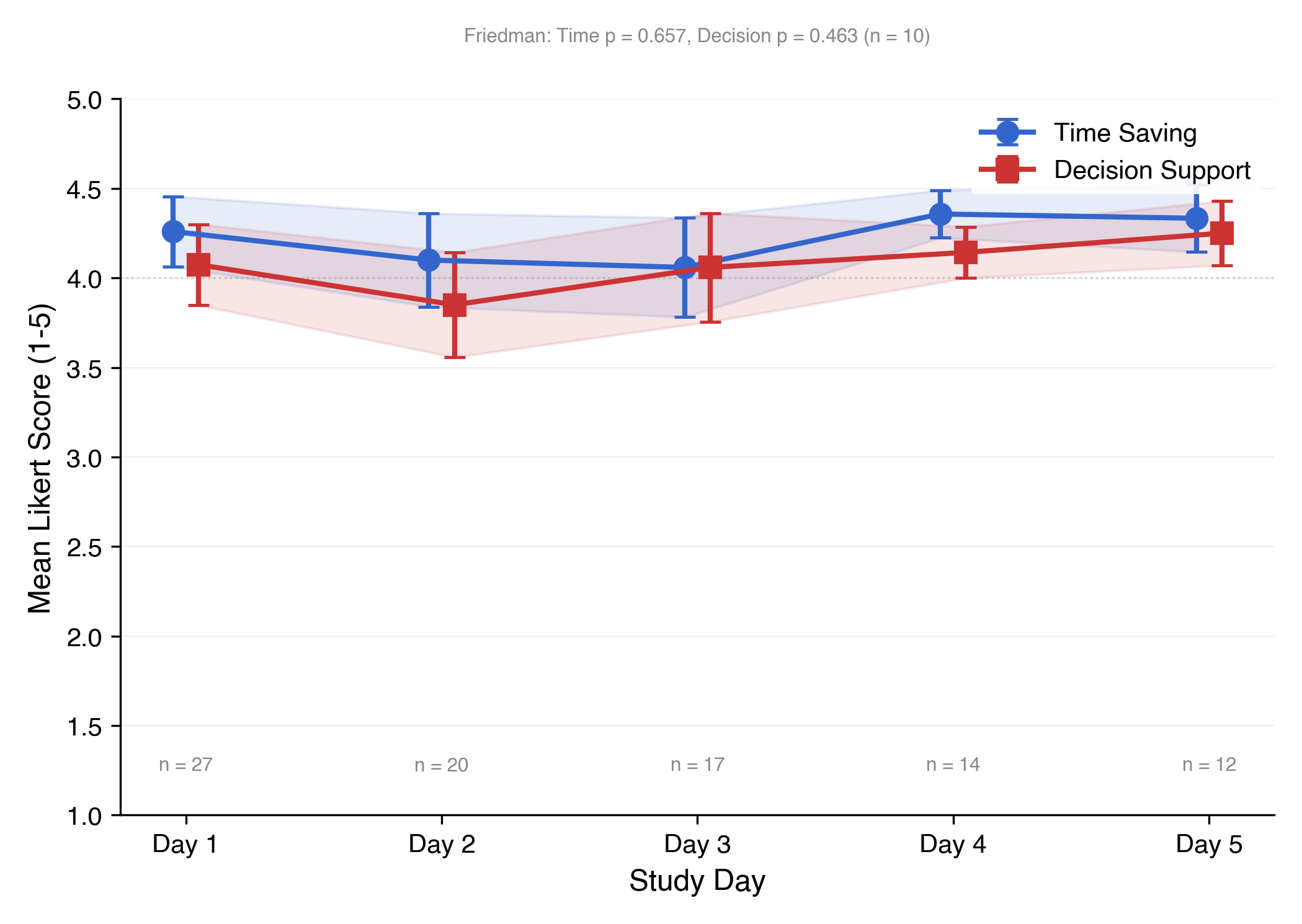}
    \caption{Mean Likert scores for time saving and decision support across the five study days. Mean daily Likert scores for physician-perceived time saving and decision support plotted across the five study days. Error bars represent standard error of the mean. The number of respondents decreased from 27 on day one to 12 on day five. The Friedman test, applied to the 10 participants with complete five-day data, showed no significant change over time for either time saving (p = 0.657) or decision support (p = 0.463).}
    \label{fig:longitudinal}
\end{figure}

No significant temporal trend was detected across the five study days. Among the 10 participants who completed all five days, the Friedman test showed no significant change over time for either time saving ($\chi^2$ = 2.43, p = 0.657) or decision support ($\chi^2$ = 3.60, p = 0.463). Table~\ref{tab:longitudinal} presents the day-by-day scores, and Figure~\ref{fig:longitudinal} illustrates the trend across the study period.

\begin{table}[H]
\centering
\caption{Day-by-day scores across the study period. n refers to the number of participants with diary entries on that day. Friedman test (n = 10 participants with all five days): time saving $\chi^2$ = 2.43, p = 0.657; decision support $\chi^2$ = 3.60, p = 0.463. SD = standard deviation; SEM = standard error of the mean.}
\label{tab:longitudinal}
\begin{tabular}{lccccc}
\toprule
\textbf{Day} & \textbf{n} & \textbf{Time, mean $\pm$ SD} & \textbf{SEM} & \textbf{Decision, mean $\pm$ SD} & \textbf{SEM} \\
\midrule
1 & 27 & 4.26 $\pm$ 1.02 & 0.20 & 4.07 $\pm$ 1.17 & 0.23 \\
2 & 20 & 4.10 $\pm$ 1.17 & 0.26 & 3.85 $\pm$ 1.31 & 0.29 \\
3 & 17 & 4.06 $\pm$ 1.14 & 0.28 & 4.06 $\pm$ 1.25 & 0.30 \\
4 & 14 & 4.36 $\pm$ 0.50 & 0.13 & 4.14 $\pm$ 0.53 & 0.14 \\
5 & 12 & 4.33 $\pm$ 0.65 & 0.19 & 4.25 $\pm$ 0.62 & 0.18 \\
\bottomrule
\end{tabular}
\end{table}

\begin{figure}[H]
    \centering
    \includegraphics[width=0.95\textwidth]{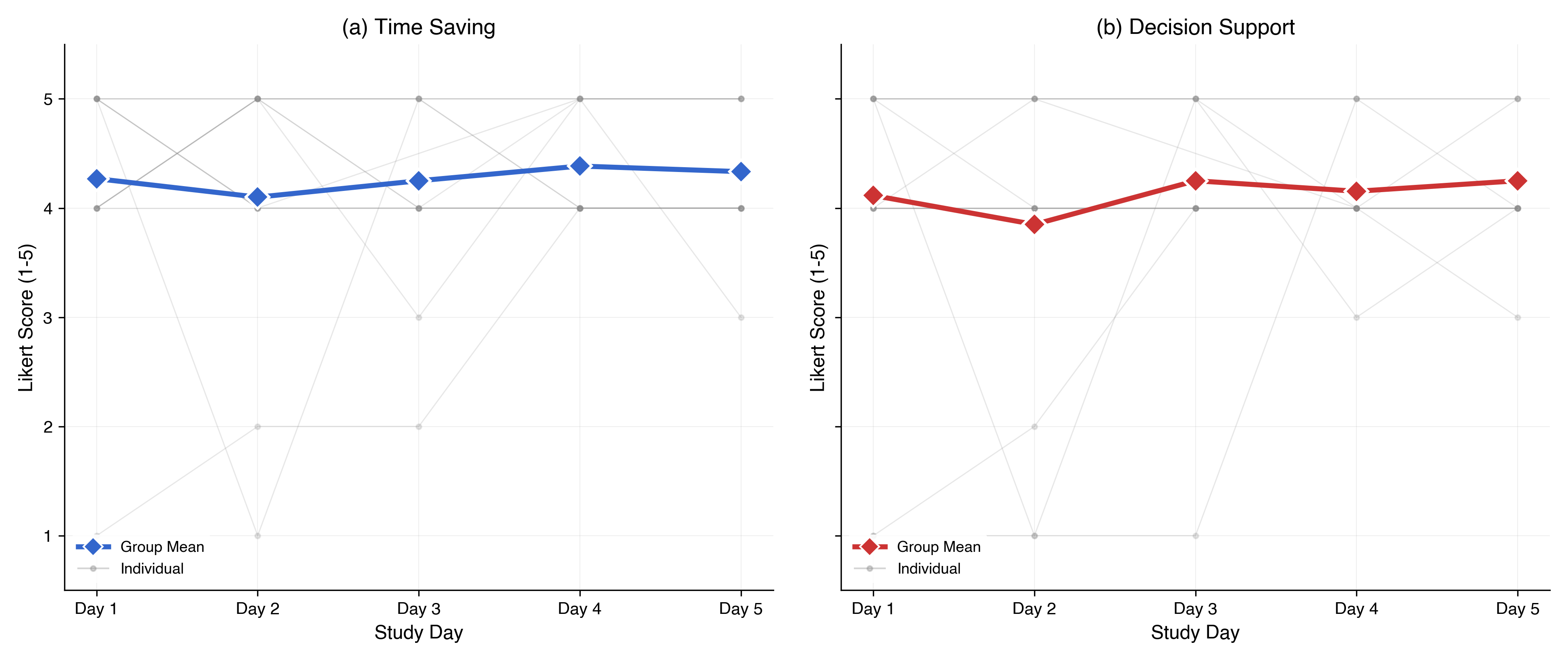}
    \caption{Heatmap of individual daily Likert scores for time saving (left) and decision support (right). Individual-level Likert scores for all 28 participants across the five study days. Rows represent participants, sorted by mean score from the highest (top) to the lowest (bottom). Columns represent study days. Green cells indicate agreement (scores 4--5), yellow indicates neutral responses (score 3), and red indicates disagreement (scores 1--2). Grey cells denote days on which no diary entry was submitted. The predominance of green across both panels is consistent with the high aggregate scores reported in Table~\ref{tab:primary}.}
    \label{fig:trajectories}
\end{figure}

Figure~\ref{fig:trajectories} presents individual daily Likert scores for all 28 participants. The predominance of high scores across the heatmap is consistent with the aggregate findings. Grey cells denote days on which no diary entry was submitted.

\subsection{Variable analysis and correlations}

\begin{figure}[H]
    \centering
    \includegraphics[width=0.95\textwidth]{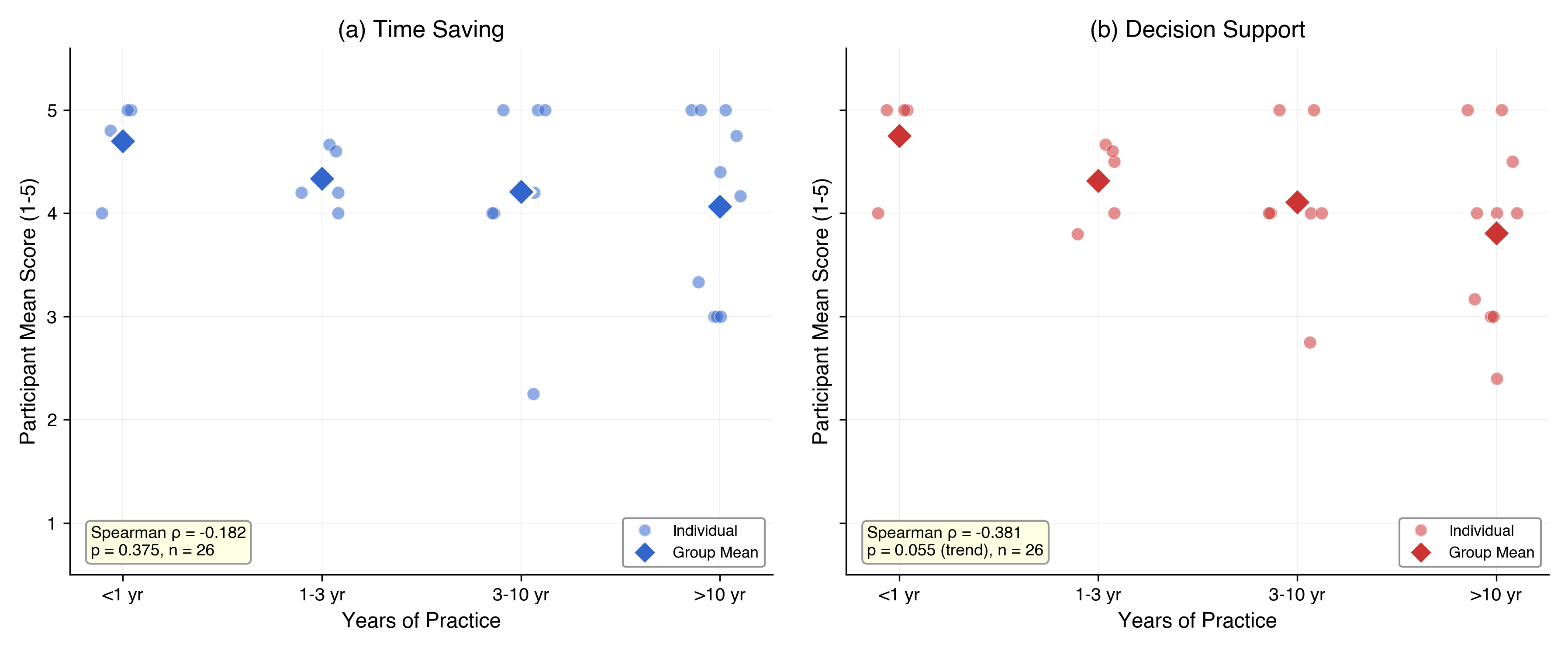}
    \caption{Scatter plots showing the relationship between years of practice and perceived value. Participant-level mean scores for time saving (left) and decision support (right) plotted against years of practice category. Each point represents one participant; larger diamond markers indicate the group mean for each experience category. Spearman's rank correlation coefficients are annotated on each panel. The association between years of practice and perceived decision support showed a moderate inverse correlation that did not reach statistical significance ($\rho$ = $-$0.381, p = 0.055, n = 26). The corresponding correlation with time saving was weaker and non-significant ($\rho$ = $-$0.182, p = 0.375).}
    \label{fig:scatter}
\end{figure}

Associations between participant characteristics and physician-perceived utility were examined using Spearman's rank correlation coefficient. An inverse correlation between years of practice and physician-perceived decision support did not reach statistical significance ($\rho$ = $-$0.381, p = 0.055, n = 26); the moderate effect size suggests a trend warranting investigation in larger samples. The corresponding correlation with time saving was weaker ($\rho$ = $-$0.182, p = 0.375). Junior physicians and students tended to report higher perceived benefit, consistent with a trend suggested by the moderate effect size, though this did not reach statistical significance and requires confirmation in larger samples. These associations are illustrated in Figure~\ref{fig:scatter}.

No association was found between self-reported technology comfort and physician-perceived time saving ($\rho$ = 0.060, p = 0.771, n = 26), suggesting that the interface was accessible regardless of prior technology proficiency.

\subsection{Net Promoter Score and general satisfaction}

\begin{figure}[H]
    \centering
    \includegraphics[width=0.95\textwidth]{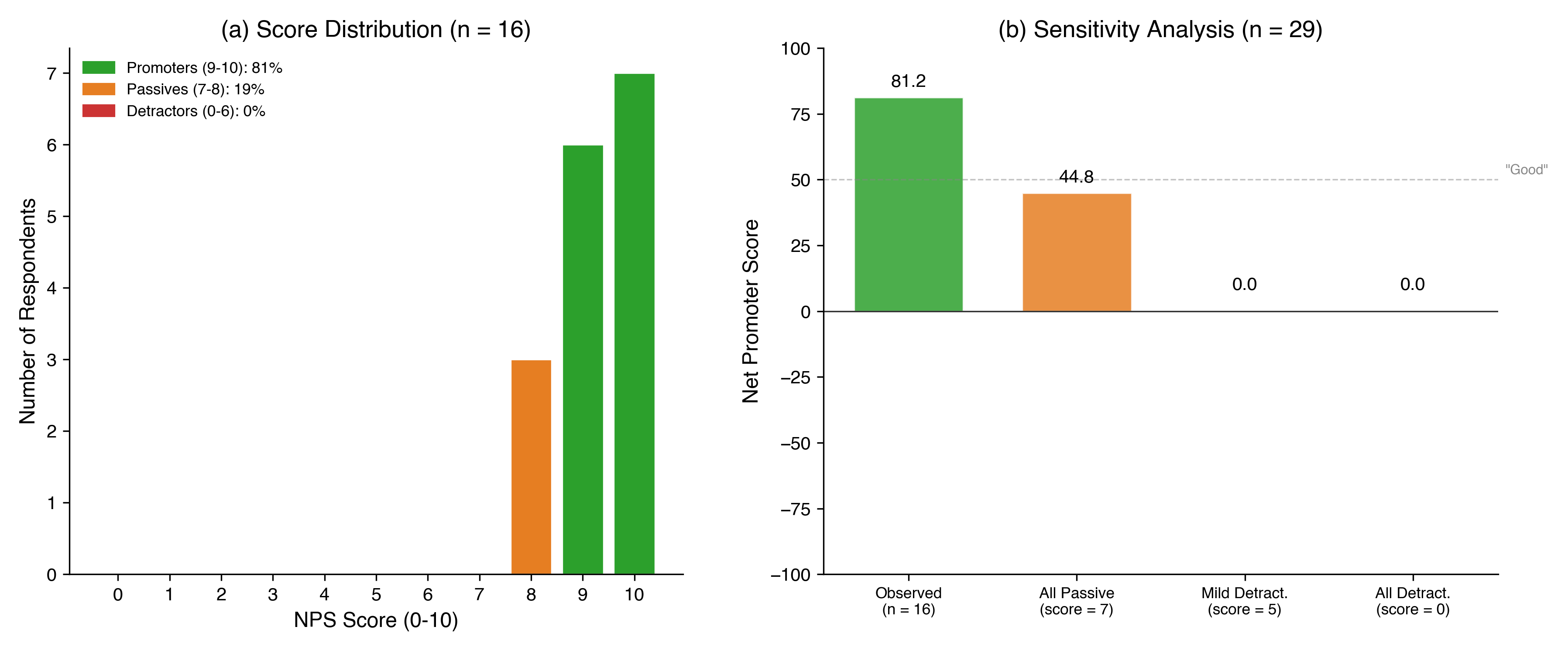}
    \caption{Distribution of NPS scores and sensitivity analysis. (a) Distribution of NPS responses among the 16 participants who completed the final evaluation. Scores of 9--10 are classified as Promoters (81\%), 7--8 as Passives (19\%), and 0--6 as Detractors (0\%). No detractors were identified. (b) Sensitivity analysis showing the resulting NPS under four assumptions for the 13 non-respondents: observed data only (NPS = 81.2), all assigned as Passives (NPS = 44.8), all assigned as mild Detractors (score = 5) (NPS = 0.0), and all assigned as strong Detractors (score = 0) (NPS = 0.0). As all scores in the Detractor range (0--6) produce the same NPS classification, both Detractor scenarios yield an NPS of 0.0. These are combined as a single scenario in Table~\ref{tab:nps_sens}. NPS = Net Promoter Score.}
    \label{fig:nps}
\end{figure}

Among the 16 participants who completed the final evaluation, DR.~INFO achieved an NPS of 81.2 (95\% CI = 62.5--100.0), with 81\% classified as Promoters (scores 9 to 10), 19\% as Passives (scores 7 to 8), and no Detractors (scores 0 to 6) \cite{nps2025}. For general reference, industry benchmarks place the average NPS for healthcare and technology sectors between 30 and 50 \cite{nps2025}, but direct comparison is not warranted given the differences in sample size and selection. Figure~\ref{fig:nps} shows the distribution of NPS scores and the sensitivity analysis results.

However, given that only 55\% of participants completed the final form, we conducted a sensitivity analysis to assess the robustness of this finding. The 13 non-respondents with diary data were assigned hypothetical scores under two increasingly conservative scenarios. NPS classifies scores 0--6 as Detractors. Because NPS is calculated as \%Promoters $-$ \%Detractors (without weighting by score magnitude), all scores within the Detractor range (whether 0 or 5) yield the same NPS. Table~\ref{tab:nps_sens} presents the resulting NPS under each scenario.

\begin{table}[H]
\centering
\caption{NPS sensitivity analysis under different assumptions for non-respondents (n = 13 with diary data). Observed NPS based on 16 respondents who completed the final eCRF. Sensitivity scenarios applied to the 13 non-respondents who submitted diary entries but did not complete the final evaluation. *: Any score in the 0--6 range yields the same NPS classification. NPS = Net Promoter Score; eCRF = electronic Case Report Form.}
\label{tab:nps_sens}
\begin{tabular}{lcc}
\toprule
\textbf{Scenario} & \textbf{Assumed non-responder score} & \textbf{Resulting NPS} \\
\midrule
Observed (n = 16 only) & -- & 81.2 \\
All non-responders Passive & 7 (Passive range) & 44.8 \\
All non-responders Detractor & 0--6 (any Detractor score)* & 0.0 \\
\bottomrule
\end{tabular}
\end{table}

The attrition analysis provides preliminary context for interpreting the NPS in light of the 45\% non-completion of the final evaluation, though the sample size is insufficient to establish equivalence between respondents and non-respondents. Under the conservative assumption that all non-respondents were Passives, the NPS would remain positive at 44.8.

\subsection{Subgroup analysis by clinical role}

Physician-perceived scores were positive across career stages, with higher mean values observed among students and residents. Given that subgroup sizes ranged from n = 1 to n = 12, formal statistical comparison across roles was not performed \cite{siegel1988}. Table~\ref{tab:subgroups} presents the descriptive scores by clinical role.

\begin{table}[H]
\centering
\caption{Subgroup analysis of perceived value by clinical role. N = 26. Values reported as mean $\pm$ SD. Time saving and decision support scored on five-point Likert scales (1 = strongly disagree, 5 = strongly agree). SD = standard deviation.}
\label{tab:subgroups}
\small
\begin{tabular}{lccc p{4.5cm}}
\toprule
\textbf{Role} & \textbf{n} & \textbf{Time saving} & \textbf{Decision support} & \textbf{Primary use cases} \\
\midrule
Medical Student & 4 & 4.70 $\pm$ 0.48 & 4.75 $\pm$ 0.50 & Individual study, case preparation \\
Resident & 6 & 4.44 $\pm$ 0.37 & 4.43 $\pm$ 0.45 & Guidelines, drug dosing, differentials \\
Attending Physician & 12 & 4.03 $\pm$ 0.99 & 3.94 $\pm$ 0.72 & Complex cases, multimorbidity \\
Consultant & 3 & 4.11 $\pm$ 0.84 & 4.06 $\pm$ 0.92 & Teaching, case review, infographics \\
Head of Service & 1 & 4.40 & 2.40 & Decision support, bibliography \\
\bottomrule
\end{tabular}
\end{table}

One Head of Service participant rated decision support at 2.40/5 while rating time saving at 4.40/5, but this represents a single participant and cannot be generalized.

\subsection{Clinical use case patterns}

Reported use cases spanned a broad range of clinical scenarios. The most frequently selected categories were treatment advice (n = 47), disease details and differential diagnosis (n = 46), medication details (n = 36), diagnosis (n = 36), and individual study (n = 31), the latter primarily reported by medical students and residents. Figure~\ref{fig:usecases} presents the full distribution of use case categories.

\begin{figure}[H]
    \centering
    \includegraphics[width=0.8\textwidth]{plots/fig7_use_cases.png}
    \caption{Frequency of clinical use case categories reported by participants across all diary entries. Frequency with which each clinical use case category was selected across all 90 diary entries. Treatment advice (n = 47) and disease details and differential diagnosis (n = 46) were the most commonly reported use cases, followed by medication details (n = 36), diagnosis (n = 36), and individual study (n = 31). Less frequently reported categories included consultation with patients (n = 22), exam preparation (n = 16), and emergency room use (n = 13). Participants could select multiple categories per diary entry.}
    \label{fig:usecases}
\end{figure}

\subsection{Qualitative feedback analysis}

\begin{figure}[H]
    \centering
    \includegraphics[width=0.8\textwidth]{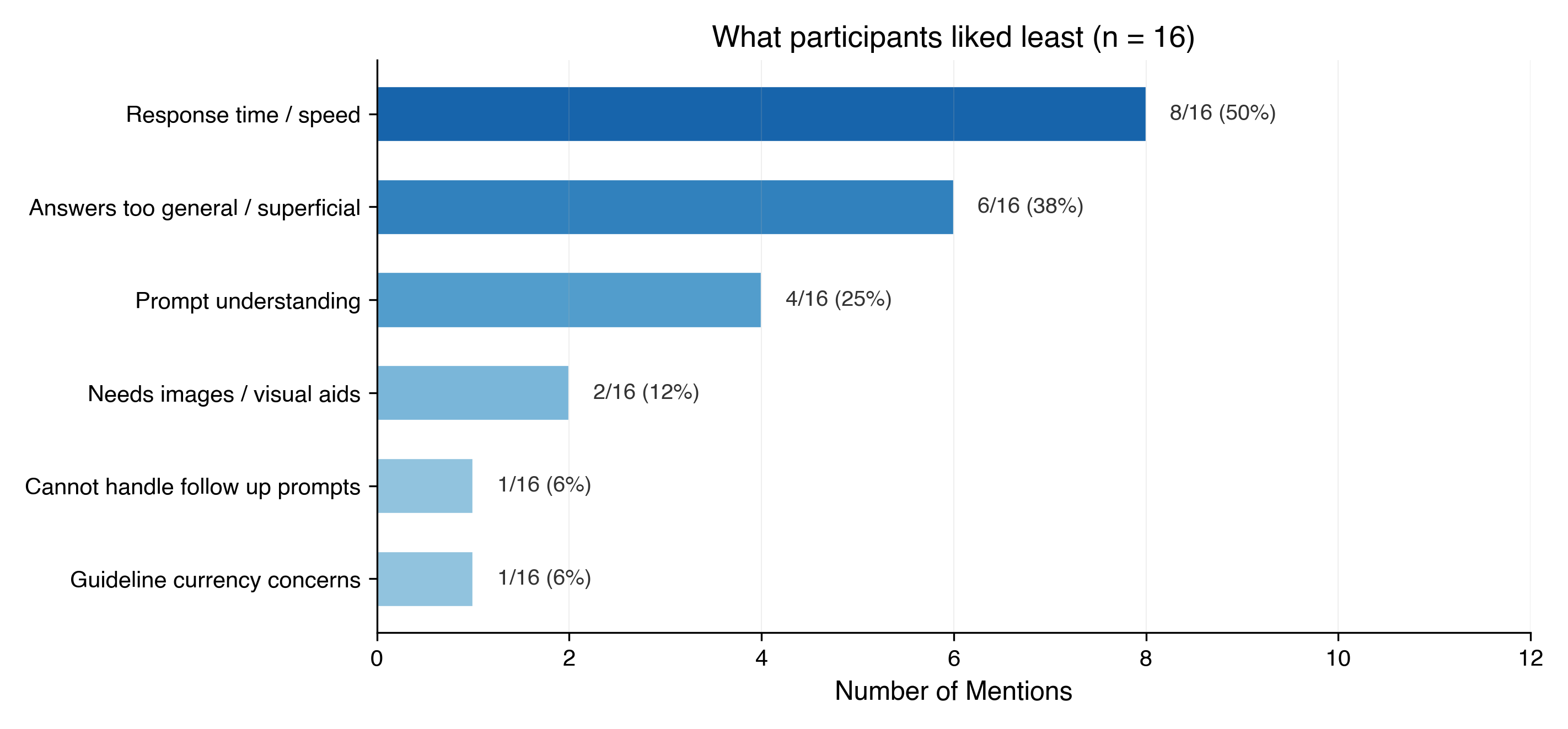}
    \caption{Content analysis of participant feedback on areas for improvement. Frequency of themes identified from content analysis of the 16 open-ended responses to the question ``What did you like least about DR.~INFO? What could be improved?'' Response time and speed were the most frequently cited concerns (8/16, 50\%), followed by answers being too general or superficial (6/16, 38\%), prompt understanding limitations (4/16, 25\%), need for images and visual aids (2/16, 12\%), inability to handle follow-up prompts (1/16, 6\%), and guideline currency concerns (1/16, 6\%).}
    \label{fig:themes}
\end{figure}

Content analysis of the 16 open-ended responses regarding areas for improvement identified five recurrent themes. The most frequently cited concern was response latency (50\%), followed by insufficient specificity of answers (38\%) and limitations in understanding nuanced clinical queries (25\%). Figure~\ref{fig:themes} presents the full breakdown of these themes.

\section{Discussion}

Several features of the study population provide important context for interpreting these findings. First, 96\% of the 26 participants with baseline data were already frequent users of established clinical decision-support tools (e.g., UpToDate, Amboss, BMJ Best Practice). The high ratings, therefore, came from participants accustomed to such resources, not from users encountering AI-assisted tools for the first time.

Second, the mean time-saving score of 4.27/5 (95\% CI = 3.97--4.57) was reported by physicians who already had established workflows with comparable tools. This study did not directly compare DR.~INFO against specific alternatives; the ratings reflect perceived utility rather than measured incremental performance. A recent evaluation of OpenEvidence, another AI clinical decision-support platform, reported physician satisfaction scores of 3.60 $\pm$ 0.38 out of 4 across common primary care conditions, with the platform primarily reinforcing rather than modifying clinical plans \cite{hurt2025}.

Third, no participant reported using clinical decision-support tools less than weekly, indicating that the cohort was well-positioned to evaluate whether an AI clinical assistant adds practical value to an already digitally supported workflow.

Technology comfort was not associated with perceived time saving (Spearman $\rho$ = 0.060, p = 0.771). Satisfaction scores remained stable across the five-day study window (Friedman test: time saving p = 0.657, decision support p = 0.463), though the short duration and low statistical power (n = 10) limit this observation. This stability should not be interpreted as evidence that a novelty effect can be excluded over longer timeframes.

Response latency was the most frequently cited concern (50\% of respondents). This reflects the additional processing time required when the system retrieves and verifies information from multiple sources before generating a response. Subsequent development iterations have targeted response time reduction. Perceived value was observed across diverse clinical contexts, from chronic disease management in primary care to acute hospital scenarios. A scoping review of AI-based clinical decision-support systems in primary care found promising outcomes for clinical management and workload reduction, though implementation remained influenced by physician perceptions and cultural context \cite{gomezcabello2024}. Given that several role-based subgroups contained fewer than five participants, these data are presented descriptively (Table~\ref{tab:subgroups}) rather than through formal statistical comparison. The distribution of participant-level means (interquartile range (IQR) = 4.0--5.0 for both metrics) suggested that positive ratings were broadly distributed rather than driven by a small number of outliers. Together with the absence of an association between technology comfort and perceived utility ($\rho$ = 0.060, p = 0.771), these findings suggest that the tool was accessible and useful across career stages and technical backgrounds.

Beyond latency, 38\% of respondents cited insufficient specificity of answers, and 25\% noted limitations in understanding complex clinical queries. These areas have been prioritized for subsequent development iterations. High overall satisfaction scores alongside specific criticisms in the open-ended responses are not contradictory. The Likert items assessed general satisfaction, whereas the open-ended question directly asked participants what they liked least, naturally eliciting improvement suggestions even from satisfied users.

\subsection{Attrition and potential bias}

With only 55\% of participants completing the final evaluation, selective dropout may have biased the results. The attrition analysis showed comparable diary scores between completers and non-completers (p = 0.672 and p = 0.886), and the NPS sensitivity analysis yielded a score of 44.8 when all non-respondents were assumed to be Passives, and 0.0 when assumed to be Detractors (Table~\ref{tab:nps_sens}). These observations suggest the two groups were similar, but the sample size does not allow formal equivalence testing, and selective dropout cannot be excluded.

\subsection{Broader implications: time efficiency and cognitive load}

Although this study did not directly measure burnout or administrative workload, the high proportion of entries reporting perceived time efficiency (87.8\%) may have implications for clinical workflow. Physician burnout has been consistently linked to administrative burden and cognitive overload \cite{berg2025,shanafelt2025}, and emerging evidence suggests that AI-based clinical assistants may help reduce information-seeking time \cite{oreilly2025}. Causal relationships cannot be inferred from this pilot, and future studies should incorporate validated burnout and cognitive load instruments to examine these outcomes more rigorously.

The inverse correlation between seniority and perceived decision support ($\rho$ = $-$0.381, p = 0.055), although not statistically significant, suggests that the tool may be particularly useful in medical education and residency training, where rapid access to guideline-based information directly supports learning. A systematic review encompassing 60 studies across 39 countries found that over 60\% of physicians and medical students reported positive attitudes toward clinical AI, with younger physicians and medical students generally expressing greater openness \cite{chen2022}. For senior physicians, the perceived value appears to center on verification and rapid cross-referencing against current evidence. Notably, treatment-related queries were the most frequently reported use case (Figure~\ref{fig:usecases}), consistent with the clinical orientation of the cohort. This pattern highlights why prospective accuracy verification of AI-generated clinical content is needed.


\section{Limitations}

Several limitations should be acknowledged. First, and most significantly, the study did not include independent verification of the clinical accuracy of DR.~INFO's responses. Physician satisfaction does not constitute evidence of clinical correctness, and in the context of AI hallucinations, outputs that appear credible but are factually incorrect represent a genuine patient safety concern. Although the companion HealthBench evaluation \cite{ravichandran2025} provides preliminary accuracy data across the 1,000-case Hard subset, benchmark performance does not substitute for verification of the specific responses generated during this pilot. Prospective accuracy auditing of real-world clinical queries by independent clinical experts is a necessary next step.

Second, this was a single-arm pilot study without a comparator group, and all outcomes reflect physician perceptions captured via self-administered Likert scales. The findings cannot be attributed specifically to DR.~INFO versus contextual factors, and social desirability or expectancy effects cannot be excluded. The five-day observation period was sufficient to assess initial acceptability, but it cannot rule out novelty effects over longer timeframes.

Third, the use of five-point Likert scales, combined with high baseline satisfaction, may have produced ceiling effects that limit the ability to distinguish between degrees of positive perception. Future studies should consider wider response scales.

Fourth, all participants were recruited from Portuguese healthcare institutions, and the sample consisted predominantly of physicians already experienced with clinical decision-support tools. The small sample size (29 participants), single-country recruitment, and high baseline digital adoption constrain the generalizability of these findings.


\section{Conclusions}

In this prospective pilot study of 29 participants involving physicians and medical students, DR.~INFO was rated positively for both time efficiency and clinical decision support, with scores stable within the five-day study window and an NPS of 81.2 among the 16 participants who completed the final evaluation. Sensitivity analysis indicated the NPS remained positive (44.8) under conservative non-response assumptions. These findings reflect physician perceptions rather than objective clinical impact and are hypothesis-generating. Ratings were consistent across clinical roles and independent of prior technology proficiency, with a non-significant trend toward higher perceived decision support among junior participants. Participant feedback identified response latency and answer specificity as the primary areas for improvement. This pilot did not include independent verification of the clinical accuracy of DR.~INFO's responses, and prospective accuracy auditing of real-world clinical queries remains a necessary next step. Larger controlled studies with objective performance measures, accuracy verification, and recruitment across multiple centers are needed to determine whether these perceived benefits lead to measurable clinical impact.


\section*{Author Contributions}

\textbf{Concept and design:} Valentine Emmanuel Gnanapragasam, Sandhanakrishnan Ravichandran, Miguel Romano, Rog\'{e}rio Corga da Silva, Shivesh Kumar, Michiel van der Heijden, Olivier Fail, Tiago Mendes, Marta Isidoro.

\textbf{Acquisition, analysis, or interpretation of data:} Valentine Emmanuel Gnanapragasam, Sandhanakrishnan Ravichandran, Miguel Romano, Rog\'{e}rio Corga da Silva, Shivesh Kumar, Tiago Mendes, Marta Isidoro.

\textbf{Drafting of the manuscript:} Valentine Emmanuel Gnanapragasam, Sandhanakrishnan Ravichandran, Miguel Romano, Rog\'{e}rio Corga da Silva, Shivesh Kumar, Michiel van der Heijden, Olivier Fail.

\textbf{Critical review of the manuscript for important intellectual content:} Valentine Emmanuel Gnanapragasam, Sandhanakrishnan Ravichandran, Miguel Romano, Rog\'{e}rio Corga da Silva, Shivesh Kumar, Michiel van der Heijden, Olivier Fail, Tiago Mendes, Marta Isidoro.

\textbf{Supervision:} Valentine Emmanuel Gnanapragasam, Sandhanakrishnan Ravichandran, Miguel Romano, Rog\'{e}rio Corga da Silva, Michiel van der Heijden, Olivier Fail, Tiago Mendes.

\section*{Disclosures}

\textbf{Human subjects:} Informed consent for treatment and open access publication was obtained or waived by all participants in this study.

\textbf{Animal subjects:} All authors have confirmed that this study did not involve animal subjects or tissue.

\textbf{Conflicts of interest:} In compliance with the ICMJE uniform disclosure form, all authors declare the following. \textbf{Payment/services info:} All authors have declared that no financial support was received from any organization for the submitted work. \textbf{Financial relationships:} Rog\'{e}rio Corga da Silva, Miguel Romano, Tiago Mendes, Marta Isidoro, Sandhanakrishnan Ravichandran, Shivesh Kumar, Michiel van der Heijden, Olivier Fail, and Valentine Emmanuel Gnanapragasam declare(s) employment from Synduct GmbH. All authors are affiliated with Synduct GmbH, the developer of DR.~INFO. Valentine Emmanuel Gnanapragasam, Olivier Fail, Michiel van der Heijden, Rog\'{e}rio Corga da Silva, and Miguel Romano are co-founders of the company, with Rog\'{e}rio Corga da Silva and Miguel Romano additionally serving as Medical Officers. Sandhanakrishnan Ravichandran and Shivesh Kumar are founding engineers, and Tiago Mendes and Marta Isidoro serve as medical advisors. \textbf{Other relationships:} All authors have declared that there are no other relationships or activities that could appear to have influenced the submitted work.

\section*{Acknowledgements}

We gratefully acknowledge the technical support provided by Sonu Kumar and Kritika Singh. We also thank all participating clinicians for their time and candid feedback throughout the study.


\bibliographystyle{unsrtnat}
\bibliography{references}

\end{document}